
\input phyzzx    

\twelvepoint

\def\dmm{{\partial}_=}
\def\dpp{{\partial}_{\ne}}

\def\a{\alpha'}
\def\eab{\epsilon_{AB}}
\def\ecd{\epsilon_{CD}}
\def\eyz{\epsilon_{YZ}}
\def\eabp{\epsilon_{A'B'}}
\def\eyzp{\epsilon_{Y'Z'}}
\def\by{\bar y}
\def\bz{\bar z}
\def\bt{\bar t}
\def\bw{\bar w}

\def\dalemb#1#2{{\vbox{\hrule height .#2pt
        \hbox{\vrule width.#2pt height#1pt \kern#1pt
                \vrule width.#2pt}
        \hrule height.#2pt}}}

\REF\HT{C.M. Hull and P.K. Townsend, Nucl. Phys. {\bf B438} (1995) 109}
\REF\Democracy{P.K. Townsend, ``$p$-brane Democracy'', in the proceedings
of the  PASCOS conference, March, 1995,  hep-th/9507048}
\REF\Pol{J. Polchinski, Phys. Rev. Lett. {\bf 75} (1995) 4724}
\REF\Kutasov{D. Kutasov, ``Orbifolds and Solitons'', EFI-95-79,
hep-th/9512145} 
\REF\HPb{P. Howe and G. Papadopoulos, Nucl. Phys. {\bf
B289} (1987) 264}  
\REF\HPa{P. Howe and G. Papadopoulos, Nucl. Phys. {\bf
B381} (1992) 360} 
\REF\Witten{E. Witten, J. Geom. Phys. {\bf 15} (1995)
215, hep-th/9410052} 
\REF\RGflow{N.D. Lambert, Nucl. Phys. {\bf B460}
(1996) 221, hep-th/9508039} 
\REF\S{A. Strominger, Nucl. Phys. {\bf B343}
(1990) 167} \REF\CHS{C.G. Callan Jr., J.A. Harvey and A. Strominger,
Nucl. Phys. {\bf B359}  (1991) 611}
\REF\CHSA{C.G. Callan Jr., J.A. Harvey and A. Strominger, Nucl. Phys. {\bf B367} 
(1991) 60}
\REF\DGHR{A. Dabholkar, G. Gibbons, J. Harvey and F. Ruiz-Ruiz, Nucl. Phys.
{\bf B340} (1990) 33}
\REF\HP{J. Hughes and J. Polchinski, Nucl. Phys. {\bf B278} (1986) 147}
\REF\HS{J. Harvey and A. Strominger, Phys. Rev. Lett. {\bf 66}
(1991) 549}
\REF\CGK{E. Corrigan, P. Goddard and A. Kent, Comm. Math. Phys. {\bf 100} (1985) 1}
\REF\loops{N.D. Lambert, Nucl. Phys. {\bf B469} (1996) 68, hep-th/9510130} 
\REF\HPc{P. Howe and G. Papadopoulos, ``Twistor Spaces for HKT
Manifolds'' hep-th/9602108} 
\REF\CP{G. Curci and G. Paffuti, Nucl. Phys.
{\bf B286} (1987) 399} 
\REF\JJR{I. Jack,D.R.T. Jones and D.A. Ross, Nucl.
Phys. {\bf B307} (1988) 531} 
\REF\Ward{R.S. Ward, Nucl. Phys.{\bf B236} (1984) 381}
\REF\CDFN{E. Corrigan, C. Devchand, D.B. Farlie and J. Nuyts, Nucl. Phys. {\bf
B214} (1983) 452}  
\REF\FN{D.B. Fairlie and J. Nuyts, J. Phys. {\bf A 17}
(1984) 2867} \REF\fn{S. Fubini and H. Nicolai, Phys. Lett. {\bf 155B} (1985) 369}
\REF\GS{A. Galperin and E. Sokatchev, Class. Quan. Grav. {\bf 13} (1996) 161} 
\REF\WittenA{E. Witten, ``Some Comments on String Dynamics'', in the
proceedings of {\it Stings '95}, USC, March 1995, hep-th/9507121}
\REF\AGF{L. Alvarez-Gaum\'e and D. Freedman, Comm. Math. Phys. {\bf 91} (1983) 87}
\REF\PTb{G. Papadopoulos and P.K. Townsend, Class. Quan. Grav. {\bf 11} (1994) 515}
\REF\Ts{G. Horowitz and A.A. Tseytlin, Phys. Rev. {\bf D51} (1995) 2896}
\REF\Khuri{R.R. Khuri, Phys. Rev. {\bf D 48} (1992) 2947} 
\REF\GKT{J. Gauntlett, D. Kastor and J. Traschen, ``Overlapping Branes in
M- Theor'', hep-th/9604179}
\REF\M{G. Papadopoulos and P.K. Townsend, ``Intersecting M-branes'', \
hep-th/9603087} 
\REF\Ts{A.A. Tseytlin, ``Harmonic Superpositions of
M-branes'', hep-th/9604035 }

\pubnum={DAMTP R/96/24\cr hep-th/9605010}

\titlepage

\title{\bf Heterotic $p$-branes from Massive Sigma Models}

\centerline{N.D. Lambert\foot{nl10000@damtp.cam.ac.uk}}

\address{D.A.M.T.P., Silver Street\break
         University of Cambridge\break
         Cambridge, CB3 9EW\break
         England}

\vfil

\abstract

We explicitly construct massive $(0,4)$ supersymmetric ADHM
sigma models which have heterotic $p$-brane solitons as their conformal fixed points.
These yield the familiar gauge 5-brane and a new 1-brane solution which preserve $1/2$ 
and $1/4$ of the spacetime supersymmetry respectively. We also discuss an analogous 
construction for the type II NS-NS $p$-branes using $(4,4)$ supersymmetric models.



\endpage

\chapter{Introduction}

Constructed as
Bogomol'nyi solitons of supergravity
theories, $p$-branes have been studied extensively throughout the history of 
superstring theory. The extent of their importance is however, only now becoming
fully appreciated. The U-duality conjecture [\HT], along with other
string/string and even string/$p$-brane dualities, requires that $p$-brane soliton states
are on an equal footing as string states in the full non-perturbative
quantum theory [\Democracy]. This interpretation has led to a
fundamentally new understanding of string theory in which
$p$-branes are of central importance.

$p$-branes are associated with $(p+2)$-form fields strengths tensors (or their 10
dimensional
$(10-p)$-form Poincar\'e duals) of superstring theory. In the type II theories they
divide themselves into R-R $p$-branes: $p=2,4,6,8$
for the IIA string and $p=1,3,5,7$ for the IIB string and NS-NS $p$-branes:
$p=1,5$ for both the IIA and IIB strings. The heterotic strings contain a 
1-brane and a 5-brane in the NS sector and the type I string has the same spectrum of
$p$-branes but in the R-R sector. R-R string states have evaded the standard
methods of string theory since they do not couple naturally to the worldsheet, whereas NS
(NS-NS) states appear as the moduli of the underlying conformal sigma model. Recently
however, Polchinski has provided the R-R
$p$-brane solitons with a string theory interpretation as spacetime field
configurations around a D-brane [\Pol]. Thus the R-R $p$-branes can be
understood within the language and tools of conformal field theory, with D-branes
interpreted as a non-perturbative worldsheet configuration. On the other
hand the NS (NS-NS) $p$-branes do correspond to exact conformal field
theories [\Kutasov] (indeed we shall construct some of them here) and are
therefore much better understood within string perturbation theory.
Nevertheless the power of D-branes has led  to a wealth of recent results
on R-R $p$-branes. Furthermore NS-NS $p$-branes have also been well
studied, in part due to the extended supersymmetry which the type II
strings possess. However, heterotic $p$-branes have to date received much
less attention and this paper may be seen as a small attempt to improve
this situation.

The perturbative massless NS (NS-NS) states of string theory are
the moduli of a conformally invariant sigma model. In conformal field theory language
they correspond to a truly marginal deformation of the sigma model which does not alter
the conformal structure. The addition of a mass term into a conformal sigma model
explicitly breaks the conformal invariance. However, we may regain a conformal
theory by following the renormalization group flow into the infrared,
whereby we move into a new string vacuum - corresponding to deformation by a relevant
operator. Here we will consider massive linear supersymmetric sigma models which flow in
the infrared to conformal sigma models describing spacetime NS $p$-branes. In
particular, when the mass is set to zero these models describe ten dimensional Minkowski
spacetime, while as the mass tends to infinity (i.e. the infrared limit) they describe
the spacetime of an NS (NS-NS)
$p$-brane.\foot{In a sense these massive sigma models may themselves be thought of as 
soliton-like objects in the space of 2 dimensional quantum field theories since they
interpolate between two distinct vacua of string theory as $m$ ranges from $0$ to
$\infty$.}

From the spacetime effective action point of view extremal $p$-branes are distinguished as
they preserve some supersymmetries of the vacuum and
therefore fall into short supermultiplets. As a consequence we expect them to correspond
to true states of the full quantum theory. On the worldsheet we need to search for exact
conformal field theories. To this end we will look for sigma models with off-shell $(0,4)$
and
$(4,4)$ supersymmetry as there is a superfield argument that these theories are finite to
all orders and therefore yield exact conformal field theories [\HPb,\HPa] (there is a
technical problem with anomalies in the chiral models which will be addressed below). 

Previously Witten has constructed an $(0,4)$ supersymmetric massive
linear sigma model whose infrared fixed point describes target space ADHM instantons 
[\Witten,\RGflow]. We may tensor the resulting $(0,4)$ supersymmetric conformal field
theory with a flat sigma model on six dimensional Minkowski space to obtain the
gauge 5-brane of the heterotic string [\S,\CHS]. In this paper we will generalize
this construction to consider massive linear sigma models whose infrared fixed points are
$(0,4)$ supersymmetric sigma models on hyper-K\"ahler target spaces of dimension $4k,
k \in {\bf N}$. For
$k=1,2$ we may tensor the resulting theory with a trivial sigma model on
flat $10-4k$ dimensional Minkowski space to obtain an exact,
$c=0$ conformal field theory. Infrared fixed points with $(0,4)$ supersymmetry thus
naturally lead to a NS 5-brane and 1-brane of the heterotic string, for $k=1,2$
respectively. Similarly conformal fixed points with $(4,4)$ supersymmetry correspond to
NS-NS $p$-branes of the type II strings.

As is well known the gauge (and neutral) 5-brane solution is described by a $(0,4)$
supersymmetric sigma model which breaks half of the $N=1$, $D=10$ supersymmetry of the
spacetime effective action [\CHS]. This leads to
$(0,1)$ supersymmetry of the $D=6$
worldsheet effective action [\CHSA]. There is a similar picture for the symmetric
5-branes where the sigma model possesses $(4,4)$ supersymmetry and thus represents a
solution to the type II string as well. Again half of the N=2
D=10 spacetime supersymmetry is broken leading to $(0,2)$ or $(1,1)$ supersymmetry on the
$D=6$ worldsheet for the type IIA and IIB theories respectively. In the case of a 1-brane
there is the well known solution of Dabholkar et al. [\DGHR] which also breaks half of the
supersymmetry of the N=1 D=10 spacetime effective action. Note that as with the symmetric
5-brane, we may view the Dabholkar et al. 1-brane as both a type II and a heterotic
string soliton by embedding the gauge connection of the heterotic string into the spin
connection. However, because the corresponding sigma  model effective action is not in
`static gauge', i.e. the metric tangent to the worldsheet is
not flat, we cannot conclude that the Dabholkar et al. sigma
model effective action admits supersymmetry [\HP]. Our
construction here necessarily considers 1-branes in static
gauge and since they are explicitly constructed to admit
$(0,4)$ supersymmetry it follows that they must preserve at
least $1/4$ of the $N=1$ $D=10$ spacetime supersymmetry.
Furthermore, since this is the maximum amount of
supersymmetry a non-trivial sigma model can admit, we cannot
expect to find 1-branes in static gauge which preserve more
than $1/4$ of the spacetime supersymmetry. There also exists
in the literature another 1-brane solution, the octonionic
string [\HS]. Like the gauge 5-brane, this solution is
unique to the heterotic string.  However, the octonionic
string soliton preserves one sixteenth of the $N=1$ $D=10$
spacetime supersymmetry and therefore the worldsheet
effective action possesses only $(0,1)$ supersymmetry.

For the rest of this paper we shall attempt to find NS $p$-branes following the
above reasoning. In the next section we will construct
$(0,4)$ supersymmetric massive sigma models and calculate their
conformal fixed point. We review the case of a four dimensional target space, already
discussed [\Witten,\RGflow], leading to the gauge 5-brane. We then discuss the
case of an eight dimensional target space which has not been considered before and leads
to a new 1-brane solution of the heterotic string. In
section three we turn our attention to the type II strings, where an analogous 
construction with
$(4,4)$ superconformal fixed points leads to NS-NS
$p$-branes, although we will not present the details of these models here.

\chapter{$(0,4)$ Supersymmetry}

Here we shall use an on-shell (0,4) supersymmetric linear sigma
model first constructed in [\Witten]. The model
consists of $4k$ bosons $X^{AY}$, $A=1,2,\ Y=1,2...,2k$ with
right handed superpartners $\psi^{A'Y}_-$, $A'=1,2$. There is also a
similar multiplet of fields $\phi^{A'Y'},\ \chi_-^{AY'}$ $Y'=1,2...,2k'$. In
addition there are $n$ left handed fermions $\lambda^a_+$, $a=1,2...,n$. The
$A,B...$ and $A',B'...$ indices are raised (lowered) by the two
by two antisymmetric tensor $\epsilon^{AB}$ ($\epsilon_{AB}$),
$\epsilon^{A'B'}$ ($\epsilon_{A'B'}$). The $Y,Z...$ and $Y',Z'...$ indices are
raised (lowered) by the invariant tensor of $Sp(k)$, $Sp(k')$ respectively
which are denoted by $\epsilon^{YZ}$ ($\epsilon_{YZ}$),
$\epsilon^{Y'Z'}$ ($\epsilon_{Y'Z'}$).

The mass terms arise from a tensor $C^a_{AA'}(X,\phi)$,   
$$ 
C^a_{AA'} = M^a_{AA'} + \epsilon_{AB}N^a_{A'Y} X^{BY} + 
\epsilon_{A'B'}D^a_{AY'}\phi^{B'Y'}  + \epsilon_{AB}\epsilon_{A'B'}E^a_{YY'}
X^{BY} \phi^{B'Y'} \ , 
\eqn\onesusy
$$
where $M^a_{AA'}$,$N^a_{A'Y}$,$D^a_{AY'}$ and $E^a_{YY'}$ are constant tensors,
which must satisfy the constraint 
$$
C^a_{AA'}C^a_{BB'}+C^a_{BA'}C^a_{AB'} = 0 \ .
\eqn\twosusy
$$
The action for the theory is given by 
$$\eqalign{
S = \int\! d^2x & \left\{ \eab\eyz \dmm X^{AY} \dpp
X^{BZ} +i\eabp\eyz \psi_-^{A'Y} \dpp \psi_-^{B'Z}
\right. \cr & \left.
+\eabp\eyzp\dmm \phi^{A'Y'} \dpp \phi^{B'Z'}
+i\eab\eyzp\chi_-^{AY'} \dpp \chi_-^{BZ'}  \right. \cr
& \left.
+i\lambda_+^a \dmm \lambda^a_+
-{im\over2}\lambda_+^a
\left(\epsilon^{BD}{\partial C^a_{BB'}\over \partial X^{DY}}\psi_-^{B'Y}
+\epsilon^{B'D'}{\partial C^a_{BB'}\over \partial
\phi^{D'Y'}}\chi_-^{BY'}\right)
 \right. \cr 
& \left.
-{m^2\over8}\epsilon^{AB}\epsilon^{A'B'}C^a_{AA'}C^a_{BB'} \right\}\ , \cr}
\eqn\action
$$
where 
$$
\dpp = {1\over{\sqrt{2}}}(\partial_0 + \partial_1 )\ \ \  \ \ \ \ \ \ \  
\dmm = {1\over{\sqrt{2}}}(\partial_0 - \partial_1 )\ ,
$$ 
and $m$ is an arbitrary mass parameter. This theory then has on-shell (0,4)
supersymmetry and, as is discussed by Witten [\Witten]for $k=1$, parallels  the
ADHM construction of instantons with instanton number $k'$ in a four
dimensional target space. In this paper we will generalise this
construction to $4k$ dimensions.

Here we will consider models for which 
$k'=1$ and $n=4k+4$. The right handed fermions are $\lambda^a_+ =
(\lambda^{AY'}_+,\lambda^{YY'}_+)$ and $C^a_{AA'}$ is chosen to be 
$$
C^a_{AA'} = B^a_{AY'}\phi_{A'}^{\ Y'} \ ,
$$
with 
$$\eqalign{
B^{YY'}_{BZ'} &= 
\epsilon_{B'C'}X_B^{\ Y} \delta^{Y'}_{Z'}\ ,\cr
B^{AY'}_{BZ'} &=
{\rho\over\sqrt2}\delta^A_B\delta^{Y'}_{Z'}\ , \cr} 
\eqn\Bdef
$$
where $\rho$ is an arbitrary positive constant,
the size of the $p$-brane core, which we will assume is non-zero for now.
The bosonic potential for this theory is 
$$
V = {m^2\over8}(\rho^2 + X^2)\phi^2 \ ,
\eqn\V
$$
where $X^2 = \epsilon_{AB}\epsilon_{YZ}X^{AY}X^{BZ}$ and similarly for
$\phi^2$. Thus, for $\rho \ne 0$, the vacuum states of the theory are
defined by $\phi^{A'Y'}=0$, and parameterize ${\bf R}^{4k}$. The
$X^{AY}$ and $\psi_-^{A'Y}$ are massless fields while $\phi^{A'Y'}$ and
$\chi_-^{AY'}$ are massive. This yields exactly four of the
$\lambda^a_+$ massive and $4k$ massless. 

To find the conformal fixed point of this model we must first determine
which are the massive and massless fields. We then integrate over the
massive fields in the path integral and follow the renormalization group
flow into the infrared limit. This procedure has been discussed before
[\RGflow] and so we shall only quote the results here. Due to its $(0,4)$
supersymmetry the action \action\ is ultraviolet finite to all orders of
perturbation theory and hence the renormalization group flow is
trivial. Furthermore as the massive fields appear only quadratically
integrating over them is exact at one loop. 

We already know that $X^{AY}$ and $\psi_-^{A'Y}$ are massless while 
$\phi^{A'Y'}$ and $\chi_-^{AY'}$ are massive.  The $\lambda^a_+$ we
must split up as
$$
\lambda^a_+ = v^a_i \zeta^i_+ + u^a_I\zeta_+^I \ ,
\eqn\lzZ
$$ 
where we have introduce an orthonormal basis of zero modes of $B^a_{AY'}$
$$
v^a_iB^a_{AY'} = 0 \ \ \ \ \ \ \ v^a_iv^a_j = \delta_{ij} 
\eqn\vdef
$$
and a similarly defined orthonormal basis for the image of
$B^a_{AY'}$, $u^a_I$. The massless components of $\lambda^a_+$ are therefore 
$\zeta^i_+$ and the massive components $\zeta^I_+$. For the potential given in
\Bdef\ we obtain 
$$\eqalign{
v^{YY'}_{ZZ'}& = \left(\delta^{Y}_{Z} -
a(X^2)\eab\ecd X^{C}_{\ Z}X^{DY}\right)\delta^{Y'}_{Z'}\  , \cr
v^{AY'}_{ZZ'}& =
{\sqrt{2}\over\sqrt{\rho^2 + X^2}}X^{A}_{\ Z}\delta^{Y'}_{Z'} \ , \cr} 
\eqn\v
$$
where
$$
a(X^2) = -{2\over X^2}\left(1-{\rho\over\sqrt{\rho^2+X^2}}\right) \ .
\eqn\adef
$$
When $k=1$ we may write $\eab X^{A}_{\ Y}X^{B}_{\ Z} = {1\over2}X^2\eyz$ and these
expressions just reduce to those found in [\Witten].

Postponing the problem of chiral and supersymmetry anomalies until later the
infrared conformal fixed point is a $(0,4)$ supersymmetric sigma model with
flat target space ${\bf R}^{4k}$ and Yang-Mills connection 
$$
A_{ijAX} = v^a_i{\partial v^a_j \over \partial X^{AX}} \ ,
$$
which we obtain to be
$$
A_{YY'ZZ'AX} = 2a\eyzp\epsilon_{(Y|X}X_{A|Z)} 
+ a^2\eyzp\ecd X_{A(Y}X^{C}_{Z)}X^{D}_{\ X}
\ . 
\eqn\Adef
$$
The field strength of this connection can be determined from the standard expression
or as the four fermion term in the tree level contribution to the integral over the
massive modes [\RGflow]. Either way we obtain
$$\eqalign{
F^{TT'UU'}_{AYBZ} =
-{4\over(\rho^2+X^2)}\eab \epsilon^{T'U'}
& \left( \delta^{T}_{(Y} +a\ecd X^{CT}X^{D}_{\ (Y} \right) \cr
&\ \ \ \ \ \ \  \left( \delta^{U}_{Z)}+a\epsilon_{EF} X^{EU}X^{F}_{\ Z)} \right)  \ . \cr}
\eqn\Fdef
$$
The classical conformal fixed point action takes the form
$$\eqalign{
S=\int\! d^2x & \left\{ \eab\eyz \dpp X^{AY} \dmm X^{BZ} 
+ i\eabp\eyz \psi_-^{A'Y}\dpp\psi_-^{B'Z} \right. \cr
& \left. + i\zeta_+^{YY'}(\epsilon_{YY'}\epsilon_{ZZ'}\dmm\psi_+^{ZZ'}
+A_{YY'ZZ'AX}\dpp X^{AX}\zeta_+^{ZZ'}) \right. \cr
& \left.
-{1\over2}\zeta_{+TT'}\zeta_{+UU'}F^{TT'UU'}_{A'YB'Z}\psi_-^{A'Y}\psi_-^{B'Z}
\right\} \ , \cr} 
\eqn\iract
$$
where $F^{TT'UU'}_{A'YB'Z}$ is simply related to $F^{TT'UU'}_{AYBZ}$ by 
replacing $\eab$ with $\eabp$ in \Fdef .

In [\Witten] Witten shows that for $k=1$ the resulting field strength
is the same as that obtained using
the ADHM construction. Similarly the field strength \Fdef\ 
we constructed here corresponds to thate obtained using the
extended ADHM method [\CGK,\Ward]. Before we move on to
discuss the quantum corrections to the classical action
\iract , it is worth while making some comments on this
construction in $4k$ dimensions. Let us then temporarily
switch to a simpler, more familiar, notation. The ADHM
construction produces along wth the gauge connection 
$A_{\alpha\beta i}  = v_{\alpha}^{a}\partial
v_{\beta}^{a} / \partial X^i$, a totally antisymmetric
constant tensor $\eta_{ijkl}$ such that the associated field
strength $F_{ij}^{\alpha\beta}$ satisfies 
$$
\lambda F_{ij}^{\alpha\beta}={1\over2}\eta_{ijkl}F_{kl}^{\alpha\beta}\ .
\eqn\eigen
$$
Here $\lambda$ is an eigenvalue which we may take to be $1$ by an
appropriate redefinition of $\eta_{ijkl}$. Now because $\eta_{ijkl}$ is totally
antisymmetric it is easy to see that the standard Yang-Mills equations of
motion are satisfied by virtue of the Bianchi identity. Thus for any $k$
the extended ADHM construction produces a solution of the
$4k$ dimensional Yang-Mills equations which is ``self-dual'' in the sense of
\eigen .  In our original notation the self-duality condition \eigen\ is simply
that \Fdef\ has the form
$$
F^{TT'UU'}_{AYBZ} = \epsilon_{AB}F^{TT'UU'}_{YZ} \ ,
\eqn\Fduality
$$
where $F^{TT'UU'}_{[YZ]}=0$.

In four dimensions there is a unique choice (up to
sign) for the tensor $\eta_{ijkl}$. In this case it is well known that the
ADHM construction produces all possible self-dual field strengths. For
$k > 1$ however there is no natural choice for $\eta_{ijkl}$, and certainly
no such choice will preserve rotational $SO(4k)$ invariance.
Furthermore the ADHM construction does not produce all
the possibilities. The solution we constructed breaks the 
$SO(4k)$ rotational invariance of ${\bf R}^{4k}$ to $SU(2)
\times Sp(k)$. Note that for $k=1$,  $SU(2)\times Sp(1)
\cong SU(2)\times SU(2) \cong SO(4)$ and rotational invariance is preserved. For
$k=2$ the properties of
\Fdef\ have been studied in [\Ward] where the self-duality equations \eigen\ were
explicitly written down, although the exact form for
$F$ was not given.

Thus we obtain a non-trivial $(0,4)$ supersymmetric sigma model for the
conformal fixed point. It is important, however, to check that this model
is ultraviolet finite. While power counting arguments can be constructed
to ensure this [\HPb], they rely upon the existence of an off-shell
superspace formulation of the theory.  The original theory \action\ only
has on-shell supersymmetry but we may promote the supersymmetry of the
action \iract\  to off-shell $(0,4)$ supersymmetry using constrained
superfields [\HPb] since the gauge group is $Sp(k)$. (Note from \Fdef\
that the $SU(2)$ factor of the $Sp(k)\times SU(2)$ gauge connection  is
flat.) 

There still remains the problem of anomalies. As is well known the
chiral sigma models suffer from anomalies which may be canceled by the inclusion of a
non-trivial 3-form
$H$ such that
$$
dH = {3\over4}\a{\rm Tr} F \wedge F \ .
\eqn\Hdef
$$
However, there will be additional
anomalies as the $(0,4)$ supersymmetry is not preserved by the quantization
procedure. Indeed there is a two loop divergence for this
theory [\loops] which seems to contradict the power counting arguments. To resolve this
requires, in addition to
\Hdef , that the metric receives a correction in the form of a finite local counter term
[\HPa]. This restores the $(0,4)$ supersymmetry which is broken by the renormalization
procedure and corresponds to the field redefinition 
$$
g_{AYBZ} = \eab\eyz + \a T_{AYBZ} \ .
\eqn\Tdef
$$
In principle the antisymmetric tensor
$b_{AYBZ}$ may also require a redefinition, however this will not be needed here.
Thus the action \iract , while ultra-violet finite and therefore conformally
invariant, requires corrections in the form of finite local counter
terms.

Instead of applying the analysis of [\HPa] to eight dimensions we may find the finite
local counter term by requiring that the two loop divergences of the action \iract\  are
canceled. In this way we obtain, 
from the general calculation in [\loops],  \Tdef\ to be
$$
T_{AYBZ}={12 \over (\rho^2+X^2)}\eab\eyz
- {12\over(\rho^2+X^2)^2}\eab\ecd X^C_{\ Y} X^D_{\ Z} \ .
\eqn\T
$$ 
Such a term presumably also restores the $(0,4)$ supersymmetry although we have not
explicitly checked this (there is evidence that this must be the case [\HPc]).
$H$ can be simply determined as the Chern-Simons 3-form of the gauge connection \Adef .

In fact \T\ only ensures the $\beta$-functions vanish up to a diffeomorphism
$X^{AY}\rightarrow X^{AY} + \xi^{AY}$ of the worldsheet fields. In particular the 
metric $\beta$-function 
of the above model is actually
$$
\beta_{AYBZ} = -2\a\partial_{(AY}\xi_{BZ)} \ 
\eqn\bdef
$$
where the vector field 
$\xi^{AY}$ 
is given by the gradient of a scalar,
$$
\xi^{AY} = \partial^{AY}\varphi \ .
\eqn\vectdef
$$
The point of this discussion is that the Curci-Paffuti relations can be used to show 
that $\varphi$ is in fact the dilaton [\CP,\JJR]. To first order in
$\a$ this yields
$$
e^{2\varphi} = 1 + 6\a\left({2k\rho^2 + (2k-1)X^2 \over (\rho^2+X^2)^2}\right) \ ,
\eqn\dil
$$
where we have assumed that the dilaton vanishes at infinity.

\subsection{The 5-brane}

First we consider the $k=1$ case corresponding to a four dimensional target space.
This case has already been considered in [\Witten,\RGflow] but we may find it
by simply setting $k=1$ in the above. In this case the gauge group is
$Sp(1)\cong SU(2)$ and we obtain
$$
F^{TT'UU'}_{AYBZ}=-
{4\rho^2\over(\rho^2+X^2)^2}\eab\epsilon^{T'U'}\delta^T_{(Y}\delta^U_{Z)}\ ,
\eqn\fiveF
$$
$$
g_{AYBZ}= \left(1 + 6\a 
{2\rho^2 + X^2\over(\rho^2+X^2)^2}\right)\eab\eyz \ ,
\eqn\fiveg
$$
$$
e^{2\varphi} = 1 + 6\a\left({2\rho^2 + X^2 \over (\rho^2+X^2)^2}\right) \ .
\eqn\fiveD
$$
In four dimensions we may write $H = -\star d f$ for some function $f$. We then find from
\Hdef\ and \fiveF\ that
$$
\triangle f = {12}\a{\rho^4 \over(\rho^2+X^2)^4} \ ,
$$
which can be solved to give
$$
H = -\star d\varphi \ ,
\eqn\fiveH
$$
with $\varphi$ given in \fiveD . This is the gauge 5-brane of [\S,\CHS] to order
$\a$ and corresponds to the field  strength of an instanton on ${\bf R}^4$. It is
completely non-singular for
$\rho >0$. The solution was originally constructed as a Bogomol'nyi soliton of the
effective supergravity theory
breaking half the supersymmetry. The 5-brane has a finite 
mass per unit 5-volume [\S] which saturates the
Bogomol'nyi bound. If
we let $\rho \rightarrow 0$ we obtain the neutral 5-brane [\CHS]. The field strength
\fiveF\ vanishes (more precisely it becomes a delta function about $X^2=0$). In this case
the singularity in the metric at $X^2$ is moved an infinite spacelike distance away. Thus
the solution remains completely non-singular.

\subsection{The 1-brane}

To find a similar 1-brane solution we consider the case $k=2$, i.e. an eight
dimensional target space. The gauge field now lies in $Sp(2)\cong Spin(5)$. From \dil\
the dilaton is
$$
e^{2\varphi} = 1 + 6\a\left({4\rho^2 + 3X^2 \over (\rho^2+X^2)^2}\right) \ .
\eqn\oneD
$$
but the field strength 
\Fdef\  cannot be simplified in this case. It is instructive then to rewrite the
metric in terms of the standard complex coordinates $\{y,z,t,w\}$ on ${\bf R}^8
\cong {\bf C}^4$ 
$$\eqalign{ X^{11} &= y \ \ \ \ \ \ \ \ \ \ \ X^{22} =
\by \cr X^{12} &= z \ \ \ \ \ \ \ \ \ \ \ X^{21} = -\bz \cr X^{13} &= t \
\ \ \ \ \ \ \ \ \ \ \ X^{24} = \bt \cr X^{14} &= w \ \ \ \ \ \ \ \ \ \ \
X^{23} = -\bw \ ,\cr } \eqn\coords
$$
hence  $ X^2 = 2(y\by+z\bz+t\bt+w\bw)$. The metric is then
 $$\eqalign{
{1\over 2}ds^2 = \left(1 + {12\a \over(\rho^2+X^2)}\right)&
(\mid dy\mid^2+\mid dz\mid^2 + \mid dt\mid^2 + \mid dw\mid^2)\cr &-
{12\a\over(\rho^2+X^2)^2}(\mid d\alpha\mid^2+ \mid d\beta\mid^2)\ ,\cr}
\eqn\oneg
$$ 
where we have introduced the $Sp(2)$ invariant 1-forms
$$\eqalign{
d\alpha &= yd\by + zd\bz+td\bt+wd\bw \ ,\cr
d\beta &= ydz-zdy+tdw-wdt \ .\cr }
\eqn\oneforms
$$
The metric \oneg\ is hermitian and positive definite. The torsion $H_{AXBYCZ}$ can also
be found as the Chern-Simons 3-form,
$$
H_{AXBYCZ}= {3\a\over2}{\rm Tr}\left(A_{[AX}\partial_{BY}A_{CZ]} +
{2}A_{[AX}A_{BY}A_{CZ]}\right) \ ,
\eqn\CSA
$$ 
where the trace is over the $Sp(2) \times SU(2) $ group indices which we have
suppressed in \CSA . The exact form of $H$ is complicated and so we omit it here. 

This
solution is only invariant under the
group [\Ward]
$$
SU(2) \times Sp(2) / {\bf Z}_2 \subset SO(8) \ .
$$
While the field strength
\Fdef\ is similar to the octonionic instanton [\FN,\fn] the metric
\oneg\ is clearly different to that of the octonionic 1-brane
[\HS]. In addition the gauge group here is
$Spin(5)$, whereas the octonionic instanton is specific to an $SO(7)$ gauge group.
A further difference is that the octonionic 1-brane admits only  $(0,1)$
supersymmetry but our solution explicitly admits
$(0,4)$ supersymmetry.

We can understand the appearance of these two distinct
heterotic 1-branes from the spacetime point of view as
follows. In order for the 1-branes to preserve some
spacetime supersymmetries it is necessary to have a
self-dual field strength [\S,\CHS], i.e.
$F^{\alpha\beta}_{ij}$ is an eigenvector of \eigen . This
will ensure that the gaugino supersymmetry variation
vanishes by projecting the supersymmetry generators
$\Gamma^{ij}\epsilon$ onto a different eigenspace of
\eigen . Above four dimensions there are a variety of
meanings of ``self-duality'' corresponding to choices of the
tensor $\eta_{ijkl}$ [\Ward,\CDFN]. For example the
octonionic instanton also satisfies a set of self-duality
eqautions [\FN,\fn] but it cannot be constructed by the ADHM
method. This can be seen by noting that the ADHM
construction produces three distinct eigenspaces of \eigen\
[\CGK,\Ward] (two of which coalesce to $\lambda=-1$ when
$k=1$), whereas the octonionic instanton has only two
[\FN,\fn]. Thus as a consequence we may expect a variety
1-brane solutions of the heterotic string, corresponding to
these different notions of ``self-duality''.

Although the metric \oneg\ is asymptotically flat, its fall off in the eight transverse 
dimensions is too slow to give the 1-brane a finite ADM mass per unit length. This is a
result of the slow fall of $F$ as $X^2 \rightarrow \infty$. Since all solutions to the
Yang-Mills equations have infinite action in dimensions greater than four 
presumably all such 1-branes have infinite masses per unit length, as is
the case of the octonionic 1-brane. It is impossible to say whether or
not the solution saturates a Bogomol'nyi bound since 
$H$ falls off only as fast as $1/ X^{3}$ at transverse infinity and
therefore the Bogomol'nyi bound also diverges [\S,\CHS,\HS].

In either of the two
complex planes $\{y,z,0,0\}$ and
$\{0,0,t,w\}$ the metric \oneg\  reduces to the 5-brane metric \fiveg\ and the field
strength \Fdef\ to that of a four dimensional instanton. This suggests, along with the
fact that only $1/4$ of the spacetime supersymmetry is preserved, that this solution does
not represent a new solitonic string, but rather some kind of bound state of two gauge
5-branes, intersecting in 1-brane. Viewed as
intersecting 5-branes the infinite mass per unit length arises because
the entire energy density of the two 5-branes has been squeezed into the one dimensional
intersection.

For
$\rho>0$ the solution is completely non-singular. In  the limiting case
$\rho \rightarrow 0$ we find that the field strength tensor does not
vanish and in fact diverges at $X^2=0$, although metric again moves this
point an infinite spacelike distance away. 
If we let $\rho \rightarrow 0$ in the gauge 5-brane we obtain the neutral 5-brane and 
$H$
becomes a closed 3-form (the Yang-Mills field strength vanishes). As a result of this we
may find a
$(4,4)$ supersymmetric solution by simply setting $A_{AX}=\omega^{(-)}_{AX}$, rather 
than
$A_{AX}=0$. This yields the symmetric 5-brane [\CHS]. It is reasonable to ask
if there is a similarly related $(4,4)$ supersymmetric solution to the $(0,4)$
supersymmetric 1-brane. However, since the Yang-Mills field is non-zero
for all $\rho$ in \Fdef , $H$ is never closed and hence this solution has no 
analogue 
of the neutral 5-brane. This
is also the case for the Octonionic 1-brane. We now consider a brief
discussion of what the conformal fixed points with $(4,4)$ supersymmetry should be.

\chapter{$(4,4)$ Supersymmetry}

In order to apply an analogous analysis to the type II strings we must find
corresponding massive sigma models whose conformal fixed points admit $(4,4)$
supersymmetry. The analysis of these theories is significantly different to those above
as there are no anomalies and no extraneous two loop divergence which need to be
canceled. Indeed it may be that $(4,4)$ supersymmetry is too
restrictive to produce a non-trivial conformal sigma model in the
infrared limit. We do not address this issue here but instead we will try
to deduce the form of the conformal fixed point from general
considerations. Similar discussions  have appeared in [\GS], where fixed
points of the model \action\ with $(4,4)$ supersymmetry where studied,
and also in [\WittenA] for $k=1$ where general conformal field theory
arguments were employed.

The conditions on the potential of a linear $(4,4)$ sigma
model have been discussed in [\AGF,\PTb]. With the same bosonic fields as the $(0,4)$
supersymmetric case the potential has the form
$$
V = {1\over2}m^2 X^2 \phi^2 \ .
\eqn\IIV
$$
As before there is a rigid $SU(2)\times Sp(k)$ symmetry of the
theory. The classical vacua lie at either
$\phi^{A'Y'}=0$ or
$X^{AY}=0$ and in the
$k=1$ case there is a symmetry between the two branches. We choose the $4k$ dimensional 
vacua where
$\phi^{A'Y'}=0$. The theory then has four massive fields
$\phi^{A'Y'}$ and their fermionic partners
and $4k$ massless fields $X^{AY}$ with their fermionic
partners. The other vacuum, where $X^{AY}=0$, is four dimensional with $4k$
massive bosonic fields and their fermionic partners. Both branches of the classical
vacuum moduli space `touch' and form a singularity at the
origin $\phi^{A'Y'}=X^{AY}=0$, where the manifold structure
degenerates.  

It is not hard to convince oneself that since there is no need
for an equation of the form \lzZ , classically the resulting conformal field theory is a
trivial linear sigma model.
However, we must still consider the quantum corrections which are again exact at one
loop since \IIV\ is quadratic in the massive fields. These corrections may be expected
since there is a singularity in the vacuum moduli space at
$X^2=0$, where the fields we integrated over degenerate to zero mass. The possible
corrections to the trivial classical effective action at one loop come in the form of
finite local counter terms for the metric and torsion $g_{AYBZ}$ and $H_{CTAYBZ}$,
respectively. Following [\loops] the
counter term $T_{AYBZ}$ for $g_{AYBZ}$ must satisfy  
$$
\partial^2 T_{AYBZ} = 0 \ . \eqn\IIT
$$
The general form $T_{AYBZ}$ which satisfies \IIT\ and is invariant
under $SU(2)\times Sp(k)$ is
$$
T_{AYBZ}=\left({Q\over X^{4k-2}}+ {Q'\over
X^{4k}}\right)\epsilon_{AB}\epsilon_{YZ} -  2k{Q'\over
X^{4k+2}}\epsilon_{AB}\epsilon_{CD}X^C_{\ Y}X^D_{\ Z} \ ,
\eqn\IIT
$$
where $Q$ and $Q'$ are constants. Using the same method as above we then obtain the 
dilaton to be
$$
e^{2\varphi} = 1 + {Q\a \over X^{4k-2}} \ ,
\eqn\IIdil
$$
where we have assumed that $\varphi$ vanishes at infinity. Given this form for
$T_{AYBZ}$ the torsion counter term $X$ must satisfy
$$
\partial^{CT}X_{CTAYBZ}=0 \ .
\eqn\IIH
$$
These conditions only ensure that the sigma
model remains finite at two loops [\loops]. This is required by, but is
not by itself sufficient for, $(4,4)$ supersymmetry. In addition
$g_{AYBZ}$ and the generalized connection
$\omega^{(-)}_{AY}$ must be compatible with the hyper-K\"ahler structure 
[\HPb]. One can explicitly check that for these target space fields the
sigma model $\beta$-functions vanish to order $\a$, i.e. $R^{(-)}_{AYBZ}= {\cal
O}(\a^2)$. We will assume from our discussion that there is a choice for
$X$ which preserves the
$(4,4)$ supersymmetry. In this case the generalised curvature tensor with torsion
will satisfy the self-duality condition, analogous to
\Fduality ,  
$$
R^{(-)}_{AYBZ\ CTDU} = \epsilon_{AB}R^{(-)}_{YZ\ CTDU}
\eqn\Rduality
$$
with $R^{(-)}_{[YZ]\ CTDU}=0$ and the sigma model
$\beta$-function will vanish to all orders in $\a$.

First let us consider the case $k=1$, where the target space of the conformal
fixed point is four dimensional. Here there is a simple form for $H$, as used
above in the $(0,4)$ supersymmetric case, and we obtain the metric, dilaton and torsion
$$\eqalign{
g_{ABYZ}  &= \left( 1 + {Q \a \over {X^2}} \right) \epsilon_{AB}\epsilon_{YZ} \ ,\cr
e^{2\varphi} &= 1 + {Q\a \over X^2} \ ,\cr
H &= -\star d\varphi \ . \cr}
\eqn\IIH
$$
This is easily recognizable as the 
the symmetric 5-brane of [\CHS] and is indeed  an exact $(4,4)$ superconformal field
theory with the curvature satisfying \Rduality\ [\GS].

We now turn to the $k=2$ case. The metric, written in the same
coordinates as \oneg ,  and dilaton now take the form
$$\eqalign{
{1\over 2}ds^2 &= \left(1 + {Q\a \over X^6}+ {Q'\a \over X^8}\right)
(\mid dy\mid^2+\mid dz\mid^2 + \mid dt\mid^2 + \mid dw\mid^2)\cr 
&{\hskip 2cm}
-2{Q'\a\over X^{10}}(\mid d\alpha\mid^2+ \mid d\beta\mid^2)\ .\cr
e^{2\varphi} &= 1 + {Q\a\over X^6} \ ,\cr}
\eqn\IIoneg
$$
Unfortunately however, our simple analysis does not provide
a form for $H$. To remedy this a more detailed argument using the harmonic
superfield analysis of [\GS] would be useful.

\chapter{Comments}

In this paper we have explicitly given massive $(0,4)$ supersymmetric sigma models whose
conformal fixed points are NS $p$-branes of the heterotic string and computed the
leading order term required to cancel the anomalies. We have also considered the case
of the type II strings but not pursued the details here. We hope to fully address the 
$(4,4)$ supersymmetric sigma models in a future
work. In conclusion we would like to make some comments.

As was discussed in [\WittenA,\RGflow] for the $k=1$ $(0,4)$ model in section
two, taking the limit $\rho \rightarrow 0$ after the conformal fixed point is found yields
a non-trivial spacetime, identical to \fiveH . On the other hand, if we
start with  $\rho =0$ in \action , integrating over the massive
$\phi^{A'Y'}$ modes apparently does not provide any quantum corrections
since the connection is flat and the anomalies vanish. There is still a
singularity at $X^2=0$ where the 
$\phi^{A'Y'}$ fields degenerate to zero mass. We would then be led to the false conclusion
that the
vacuum moduli space is flat with a degenerating manifold structure at the origin. 
But this is exactly the same situation
that appears with a
$(4,4)$ theory based on the  potential
\IIV . This lends further support to the conjecture that \IIH\ and \IIoneg\ with 
$Q,Q'\ne 0$ are indeed the $(4,4)$ superconformal fixed points.

Another generalization of the above analysis would be to add
mass terms to an exact conformal sigma model with a curved target space and then
similarly obtain the infrared fixed point conformal field theory. For example the
linear model considered here could be modified to a massless $(4,4)$ supersymmetric sigma
model with a curved target space and mass terms which then  break the supersymmetry to
$(0,4)$. In particular one could consider the sigma model of the symmetric 5-brane
[\CHS]. Since this metric is conformally flat and the Yang-Mills equations conformally
invariant in four dimensions, the ADHM construction can again be applied. 

In the $k=1$ cases above it is also possible to tensor the infrared fixed point theory
with another $(0,4)$ or $(4,4)$ conformal field theory on a compact hyper-K\"ahler four
manifold $K$ (for example $K= {\bf T}^4$ or $K_3$) and then again with a trivial
sigma model on  two dimensional Minkowski space. With this construction we obtain a
1-brane in string theory compactified to six dimensions on $K$, but with the metric
and 3-form \IIH\  in the transverse space. This is the construction of [\Kutasov]
and from the ten dimensional point of view corresponds to a 5-brane wrapped around
$K$. These solutions preserve $1/2$ of the $D=6$ spacetime supersymmetry and hence
$1/2$ or $1/4$ of the $D=10$ spacetime supersymmetry for $K={\bf T}^4,K3$
respectively.

Another possibility for the $k=1$ cases is to tensor the conformal fixed
point theory with another copy of itself and then with a trivial two dimensional sigma
model. Thus we arrive at the double instanton solution [\Khuri] in the $(4,4)$
supersymmetric case and a similar heterotic version in the $(0,4)$ case. As one may
expect these solutions preserve $1/4$ of the spacetime supersymmetry. The double
instanton 1-branes have the property that the metric singularity is located at the origin
of  coordinates in either of the four dimensional target spaces and not at a single point
in the transverse space. They are therefore also most appropriately viewed as two
orthogonal intersecting 5-branes in ten dimensional spacetime [\GKT], although in this
case the metric and torsion are singular not just on the intersection of the two 5-branes
but also on the world sheets of each. Again these 1-branes have infinite masses per unit
length but this may be resolved by viewing them as intersecting 5-branes with the energy
density spread out over the two 5-branes. 

Recently there has been interest in the
intersecting $p$-branes of M theory [\M,\Ts,\GKT]. Here we have seen that
the situation is somewhat different for heterotic strings as the 1-brane
solutions discussed here do not depend on harmonic functions in the
transverse space, because the non-vanishing Yang-Mills field strength
acts as a source for Laplace's equation. In the future it would be
interesting to understand why both the double instanton and the solution
found here both occur as intersecting heterotic 5-branes and if the
octonionic string can be viewed in a similar light.

\parskip=12pt

\noindent I would like to thank G. Papadopoulos and P.K. Townsend for discussions.

\refout

\end